\documentclass[preprint2]{aastex}
\usepackage[]{natbib}
\begin{document}
\title{Intrinsic Redshifts and the Hubble Constant}
\author{M.B. Bell\altaffilmark{1} and S.P. Comeau\altaffilmark{1}}
\altaffiltext{1}{Herzberg Institute of Astrophysics,
National Research Council of Canada, 100 Sussex Drive, Ottawa,
ON, Canada K1A 0R6}

\begin{abstract}

We show that the V$_{\rm CMB}$ velocities of the Fundamental Plane (FP) clusters studied in the Hubble Key Project appear to contain the same discrete "velocities" found previously by us and by Tifft to be present in normal galaxies. Although there is a particular Hubble constant associated with our findings we make no claim that its accuracy is better than that found by the Hubble Key Project. We do conclude, however, that if intrinsic redshifts are present and are not taken into account, the Hubble constant obtained will be too high.

\end{abstract}

\section{Introduction}

If the local Universe were expanding uniformly, without peculiar velocities produced by local density perturbations, then in standard big bang cosmology the expansion of galaxies is defined by the Hubble law, $v$ = H$_{\rm o}d$, where $v$ is the recession velocity of a galaxy at a distance $d$, and H$_{\rm o}$ is the Hubble constant at the current epoch. The Hubble Key Project \citep{fre01} found a Hubble constant of H$_{\rm o} = 72 \pm 8$ km s$^{-1}$Mpc$^{-1}$.
Even an accurate local value has been difficult to define, because of the basic difficulties encountered in determining accurate distances and velocities. Although accurate redshifts are easy to measure, obtaining accurate calibrator distances and correcting for Doppler components introduced by local density perturbations is more difficult.  However, as these parameters become more accurately known, there are now other questions that need to be addressed, and the possibility that a component of the redshift may not be Doppler-related \citep{arp02,bel02d,rus02} can introduce another level of uncertainty that needs to be investigated. \citet{fre01} found consistent values for H$_{\rm o}$ in all cases except the FP clusters, however, their quoted uncertainty ($\pm8$ km s$^{-1}$ Mpc$^{-1}$) is still large. This paper examines the V$_{CMB}$ velocities of FP clusters in an attempt to see if they show evidence for discrete components that might have influenced the H$_{\rm o}$-value reported by \citet{fre01}.

\section{Analysis}

The Hubble Key Project reported consistent Hubble constants of H$_{\rm o} \sim71$ km s$^{-1}$Mpc$^{-1}$ for, a) Type Ia supernovae, b) Tulley-Fisher relation galaxies, c) surface brightness fluctuations galaxies and d) Type II supernovae. However, a much higher value of H$_{\rm o}$ = 82 km s$^{-1}$Mpc$^{-1}$ was found for fundamental plane (FP) clusters.

\begin{figure}
\hspace{-1.0cm}
\vspace{-1.0cm}
\epsscale{1.0}
\plotone{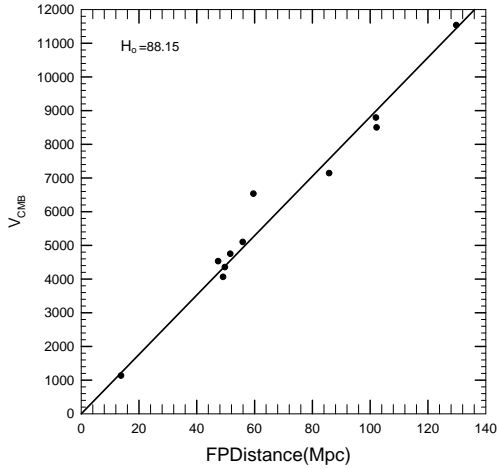}
\caption{\scriptsize{Hubble plot of Fundamental Plane clusters. Data have been taken from Table 9 of \citet{fre01}. The solid line is the best fit slope through zero velocity and gives a Hubble constant of H$_{o}$ = 88.15 km s$^{-1}$ Mpc$^{-1}$. \label{fig1}}}
\end{figure}

\begin{figure}
\hspace{-1.0cm}
\vspace{-1.0cm}
\epsscale{1.0}
\plotone{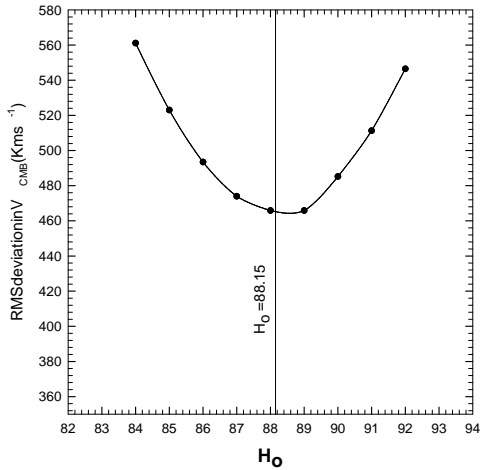}
\caption{\scriptsize{Plot of RMS deviation in V$_{\rm CMB}$ velocities about the Hubble line as a function of the Hubble constant for FP clusters listed in Table 9 of \citet{fre01}. \label{fig2}}}
\end{figure}

\begin{figure}
\hspace{-1.0cm}
\vspace{-1.0cm}
\epsscale{1.0}
\plotone{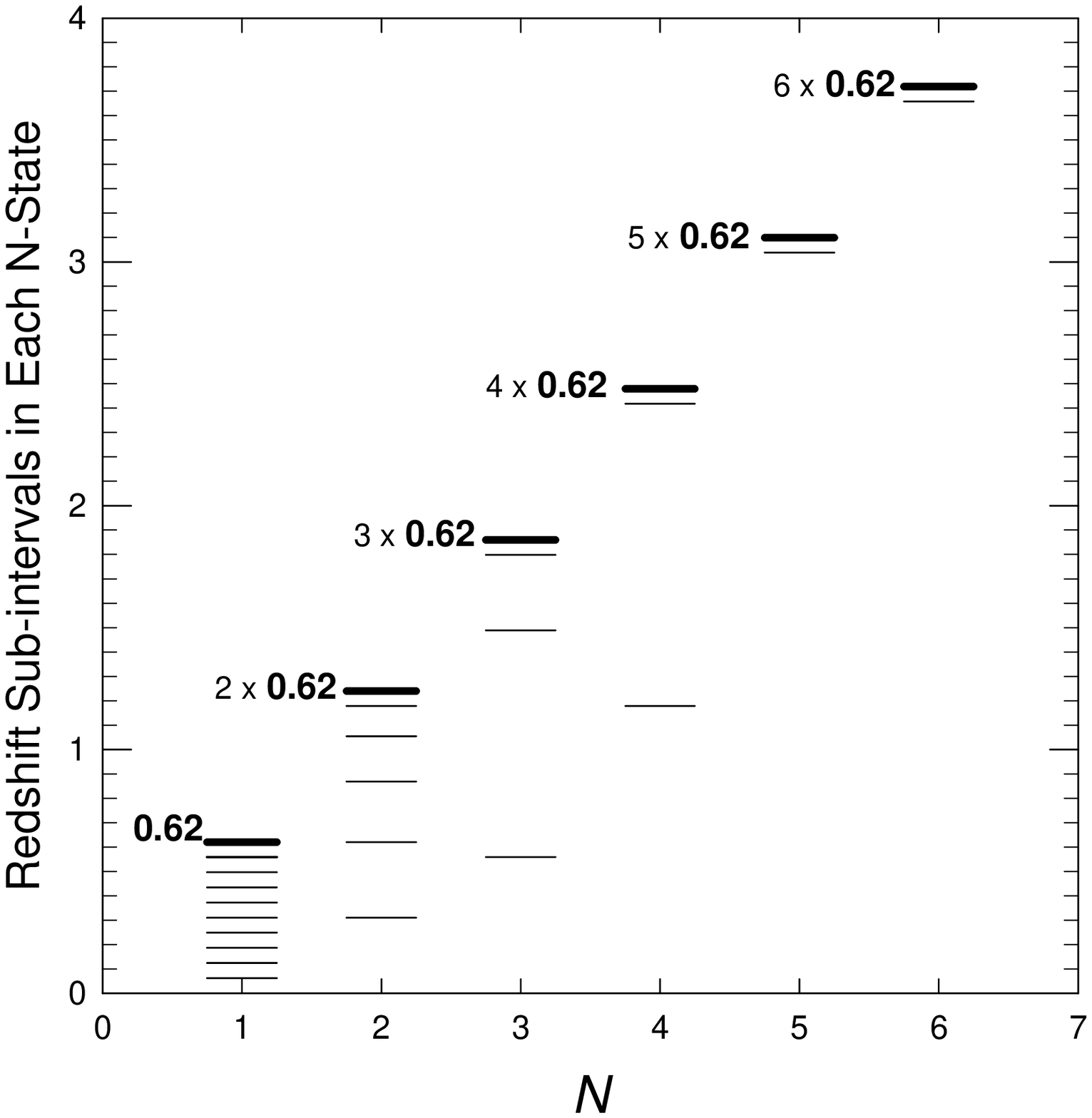}
\caption{\scriptsize{Quasar Intrinsic redshift levels for the first six N-states. \label{fig3}}}
\end{figure}

In Fig. 1, V$_{\rm CMB}$ velocities for the FP clusters listed in Table 9 of \citet{fre01} are plotted vs distance in Mpc. The line has a slope of H$_{\rm o}$ = 88.15 km s$^{-1}$ Mpc$^{-1}$ and represents the slope of a line through zero velocity and distance that gives a minimum near 465 km s$^{-1}$ in the RMS deviation in V$_{\rm CMB}$ velocities. This minimum is shown in Fig. 2 which is a plot of the RMS deviation in velocity as a function of H$_{\rm o}$. The slope found here, 88.15, is significantly higher than the value H$_{\rm o}$ = 82 reported by \citet{fre01} for the FP clusters. A linear regression calculation using the data in Fig. 1 gave a slope of 84.6 with a standard error of 4.7 km s$^{-1}$ Mpc$^{-1}$ and a zero-distance velocity intercept of 293 km s$^{-1}$.

The minimum RMS deviation in the velocities in Fig. 2 is in excess of 460 km s$^{-1}$, even though all peculiar velocities are assumed to have been removed. This value seems high, since all primordial turbulence is expected to have been damped out by adiabatic expansion \citep{kra86}. Why is the zero-distance velocity intercept (293 km s$^{-1}$) so high? Previously, peculiar velocities have been used to explain such large velocities \citep{jor96}, but this interpretation can be questioned since these large "peculiar velocities" tend to be mostly redshifts with very few blueshifts \citep{rus02}. Thus the possibility that intrinsic redshifts may play a role seems to be indicated. This suggestion is highly controversial but it may no longer be possible to ignore this possibility in light of recent evidence \citep{tif96,tif97,bel02d}.

 \subsection{Intrinsic redshifts}

A study of the compact objects near NGC 1068 \citep{bel02a,bel02b,bel02c}, and earlier work \citep{bur90}, suggested that quasar redshifts may contain an intrinsic component that is harmonically related to 0.62$\pm0.01$.
Also, \citet{tif97} was able to define several families of 'velocity' periods in normal galaxies. Recently \citet{bel02d} has pointed out that when converted to redshifts these periods are all harmonically-related to the intrinsic redshifts reported in quasars, and therefore also all harmonically tied to the redshift increment 0.62$\pm0.01$. The redshifts 0.62 and 0.062 have been shown to be important constants related to intrinsic redshifts in galaxies as well as quasars.
Evidence that 0.062 may be a significant redshift-related constant has been found in four independent investigations; first by \citet{bur68}, then by \citet{bur90}, later by \citet[and references therein]{tif97}, (although this does not seem to have been realized until pointed out by \citet{bel02d}), and finally by \citet{bel02c} using QSOs near NGC 1068. In the last instance it was recognized only after all Doppler-related redshifts were taken into accounted. 

\subsection{Intrinsic Redshift Equations}

Equations that define the intrinsic redshifts in quasars and galaxies are included in Appendix A and B respectively. Quasar intrinsic redshifts are defined by the relation z$_{\rm iQ}$[$N,n$] = z$_{f}$[$N$ - 0.1M$_{N}$], where z$_{f}$ is assumed to be a fundamental redshift constant, $N$ = 1,2,3,.., and $M_{N}$ is a function of $n$. Each z$_{\rm iQ}$ value is therefore uniquely defined by the quantum numbers $N$ and $n$. Note that eqn A1 was previously expressed in a slightly different fashion \citep{bel02d}. It has since been realized that the fundamental constant in eqn A1 is z$_{f}$ = 0.62. This can be seen clearly in Fig 3 where the allowed intrinsic redshifts for quasars are plotted for the first 6 $N$-states. z$_{f}$ = 0.62 is the fundamental constant in eqn A1 and the cornerstone relating all quasar intrinsic redshifts in Fig 3. Sub levels in each $N$-state all fall below the upper level by factors of 0.062.

Similarly, each galaxy intrinsic redshift z$_{\rm iG}$[$N,m$] is uniquely defined by the quantum numbers $N$ and $m$. The discrete redshift values found in galaxies by Tifft (expressed as velocities) were fitted to a model \citep{tif96} and the results are listed in \citet[Table 1, cols 5, 3, and 2]{tif97} for the most dominant families of periods. The corresponding discrete redshifts are listed in \citet[Table 4, cols 2, 5, and 8]{bel02d} and have been derived from eqn B1 for $N$ = 1, 2, and 3 and z$_{f}$ = 0.62. The discrete redshift components in galaxies can range from z = 0 to at least z = 0.558, however, from \citet[Table 4, cols 2,5,8]{bel02d}, most appear to be less than z$_{\rm i}$ = 0.005. Can some of these intrinsic components be present in the FP cluster velocities? If so, their presence would likely prevent the determination of an accurate Hubble constant.

\subsection{Intrinsic redshifts on a Hubble plot}

\begin{figure}
\hspace{-1.0cm}
\vspace{-2.5cm}
\epsscale{1.0}
\plotone{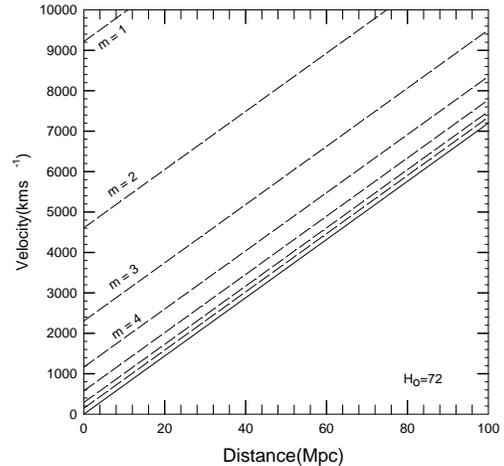}
\caption{\scriptsize{Hubble plot for normal galaxies. If the galaxies contained no intrinsic redshifts, and no peculiar velocites, then for H$_{\rm o}$ = 72, all galaxies would fall along the solid line. If their redshifts include some of the discrete intrinsic components below 10,000 km s$^{-1}$ predicted by \citet[Table 1, col 5]{tif97} and \citet[Table 4, col 2]{bel02d}, they would fall on a dashed line. To avoid confusion for small intrinsic redshift values, dashed lines representing discrete intrinsic velocities below 145 km s$^{-1}$ have not been included in the plot. \label{fig4}}}
\end{figure}

\begin{figure}
\hspace{-1.0cm}
\vspace{-1.0cm}
\epsscale{1.0}
\plotone{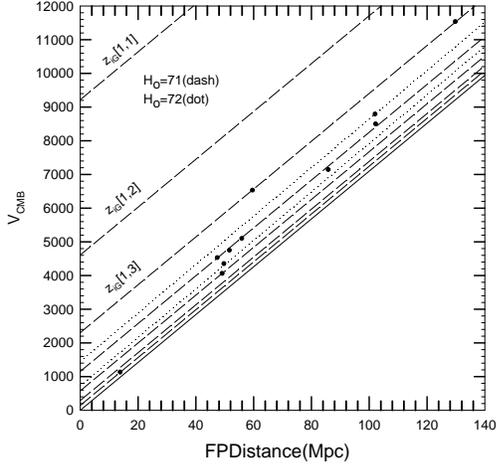}
\caption{\scriptsize{Plot of V$_{\rm CMB}$ velocities in FP clusters vs distance. Dashed lines indicate where the z$_{\rm iG}$[1,$m$] instrinsic redshift components are located for a Hubble constant of 71.0. Dotted lines are for the z$_{\rm iG}$[2,6] and z$_{\rm iG}$[2,7] intrinsic redshifts.  Data are from Table 9 of \citet{fre01}. \label{fig5}}}
\end{figure}

\begin{figure}
\hspace{-1.0cm}
\vspace{-1.0cm}
\epsscale{1.0}
\plotone{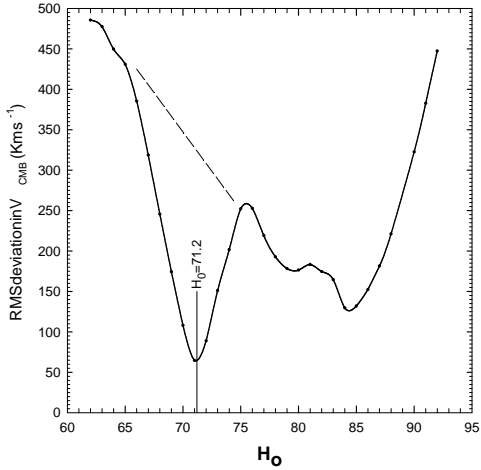}
\caption{\scriptsize{Plot of RMS deviation in V$_{\rm CMB}$ velocities for the nine $N$ = 1 sources in Fig. 4, relative to the z$_{\rm iG}$[1,$m$] ($m$ = 3,4,5,6, and 7 grid lines as a function of H$_{\rm o}$. Assumes z$_{f}$ = 0.62. \label{fig6}}}
\end{figure}

\begin{figure}
\hspace{-1.0cm}
\vspace{-1.0cm}
\epsscale{1.0}
\plotone{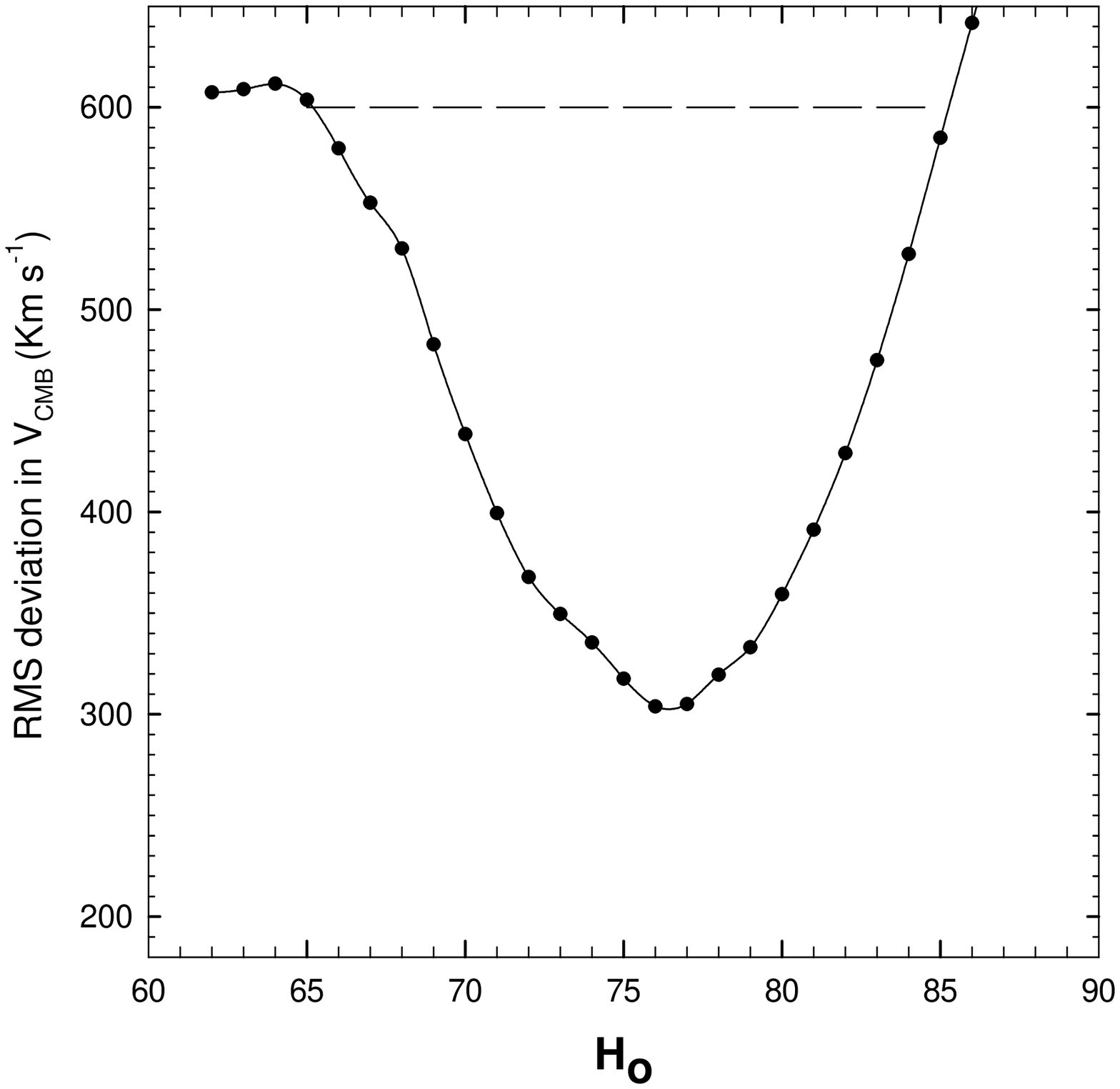}
\caption{\scriptsize{Plot of RMS deviation in V$_{\rm CMB}$ velocities relative to the z$_{\rm iG}$[1,$m$], z$_{\rm iG}$[2,6] and z$_{\rm iG}$[2,7] grid lines vs H$_{\rm o}$ for a typical test set of randomly chosen peculiar velocities as described in the text. \label{fig7}}}
\end{figure}

\begin{figure}
\hspace{-1.0cm}
\vspace{-1.0cm}
\epsscale{1.0}
\plotone{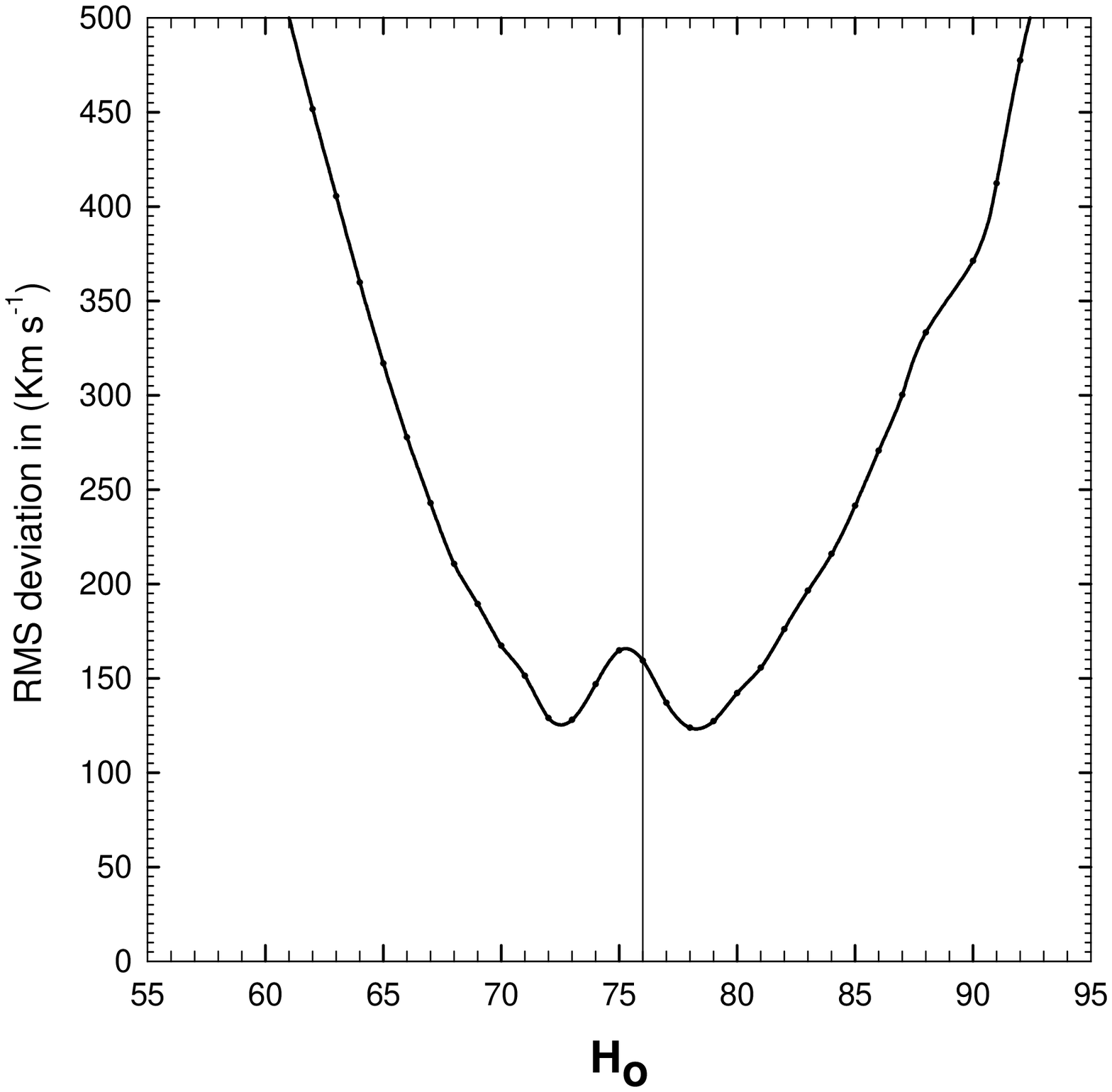}
\caption{\scriptsize{Plot of RMS deviation in V$_{\rm CMB}$ velocities relative to the z$_{\rm iG}$[1,$m$], z$_{\rm iG}$[2,6] and z$_{\rm iG}$[2,7] grid lines vs H$_{\rm o}$ for a typical test set of randomly chosen peculiar velocities as described in the text. \label{fig8}}}
\end{figure}

\begin{figure}
\hspace{-1.0cm}
\vspace{-1.0cm}
\epsscale{1.0}
\plotone{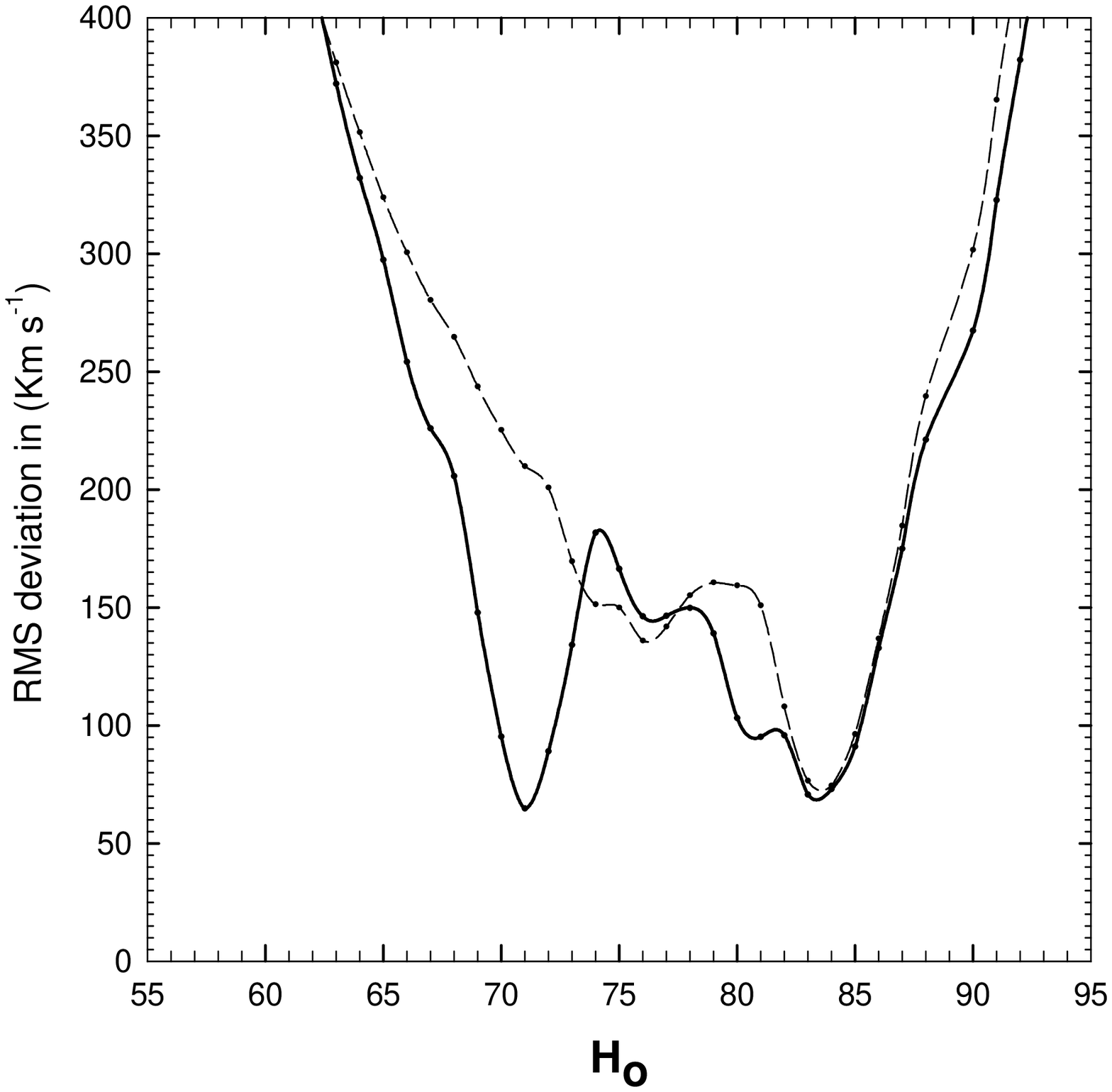}
\caption{\scriptsize{Plot of RMS deviation in V$_{\rm CMB}$ velocities relative to the z$_{\rm iG}$[1,$m$], z$_{\rm iG}$[2,6] and z$_{\rm iG}$[2,7] grid lines vs H$_{\rm o}$ for (solid curve) the FP data and (dashed curve) the same data with the peculiar velocity component polarity reversed as described in the text. \label{fig9}}}
\end{figure}

If it is assumed a) that galaxy distances are accurately known, b) that all peculiar velocities due to local density fluctuations can be accurately accounted for, and c) that the observed redshifts are entirely Doppler-related and due to the Hubble flow, (i.e. no intrinsic components present) then, if H$_{\rm o}$ = 72, all galaxies should fall along the solid line in the Hubble plot in Fig. 4. If there are intrinsic components present in some galaxies with the discrete values found by \citet{tif97} to be the most common, and defined here by eqn B1 for $N$ = 1, with $m <$ 8, then these sources will fall on the appropriate dashed lines in Fig. 4. To avoid confusion, lines for small intrinsic redshifts ($m >$ 7) have not been included in Fig. 4.

In order to test for the presence of discrete redshifts we superimposed the discrete redshift grid in Fig. 4 onto the FP data in the Hubble plot in Fig.1. The result is shown in Fig. 5 where the z$_{\rm iG}$[1,m] grid lines (dashed lines) assume H$_{\rm o}$ = 71 km s$^{-1}$ Mpc$^{-1}$ and z$_{f}$ = 0.62. The z$_{f}$-value determines the separation between grid lines while the value of H$_{\rm o}$ determines the slope. It is clear in Fig. 5 that nine of the eleven FP clusters fall along the dashed lines defined by the z$_{\rm iG}$[1,$m$] discrete redshifts with $m$ = 3,4,5,and 7. Although two of the clusters (Abell S753 and Abell 539) do not fall near dashed lines, they do fall on the two dotted lines included in the figure. These lines represent the z$_{\rm iG}$[2,6] and z$_{\rm iG}$[2,7] discrete redshifts. It is therefore assumed that these two clusters are $N$ = 2 sources while the remaining nine are $N$ = 1 sources. Fig. 5 has significance for both the model proposed by \citet{tif96,tif02} and for the scenario proposed earlier \citep{bel02b} (see section 6 below).

\subsection{RMS Deviation in V$_{\rm CMB}$ from Intrinsic Redshift Grid Lines}

In Fig. 2 we plotted the RMS deviation in V$_{\rm CMB}$ about the Hubble line in Fig 1 for a range of H$_{\rm o}$-values. It is now possible to make a similar plot of the RMS deviation in V$_{\rm CMB}$ between the sources and their nearest grid line in Fig 5, for a range of H$_{\rm o}$-values. The result is shown in Fig. 6 for the nine $N$ = 1 sources, for a grid spacing defined by z$_{f}$ = 0.62, the value found in earlier work \citep{bur90,bel02c,bel02d}. Clearly there are two places on this curve where the RMS is low, indicating a good fit to the grid lines. A broad one is located near H$_{\rm o}$ = 84 and the other, a much lower and narrower one, occurs near H$_{\rm o}$ = 71 km s$^{-1}$ Mpc$^{-1}$. The broad feature is similar to that found for the raw data in Fig. 2 and is expected. This is discussed in more detail in section 2.5 below describing randomly generated test data. The second one, near H$_{\rm o}$ = 71, corresponds to the case presented in Fig. 5 where the sources lock on to the grid lines.

\section{Test Data Using Randomly Generated Peculiar Velocities}

In order to get a feeling for what kind of results would be obtained if the velocity dispersion in Fig. 1 was due to peculiar velocities of random amplitude, instead of discrete intrinsic components, we attempted to simulate the raw FP data by generating several sets of 11 peculiar velocities. We used three different approaches to do this. 

\subsection{Test 1}

First, we generated several sets of peculiar velocities between 0 and 3500 km s$^{-1}$. We used uniform weighting and only positive values were used to resemble the real data as closely as possible. (Note that for H$_{\rm o}$ = 82, the value reported by \citet{fre01}, all peculiar velocity components except one are positive. It is not immediately obvious what weighting was used in the Hubble Key Project to obtain this value). The cutoff value of 3500 was chosen here because it falls midway between the highest intrinsic value present in the data (2314 km s$^{-1}$) and the next highest intrinsic value (4628 km s$^{-1}$). A value higher than 3500 would have been fitted to the 4628 km s$^{-1}$ level and the FP data contained no sources at this discrete velocity level or above. In each case the random-amplitude peculiar velocities we generated were added to the Hubble velocities calculated for each source using its known distance and H$_{\rm o}$ = 71 km s$^{-1}$ Mpc$^{-1}$. We then calculated the RMS deviation in V$_{\rm CMB}$ between the eleven sources and their nearest grid line, as was done above for the FP data, for a range of H$_{\rm o}$-values and z$_{f}$ = 0.62. 

All randomly generated sets gave similar results and a typical example is presented in Fig. 7. In every case the curve showed a single, broad dip in the RMS deviation, with widths covering 18-20 H$_{\rm o}$-values at a point on the curve where the RMS value was equal to twice the minimum RMS value (horizontal line in Fig. 7). The broad width and low RMS of the curve was a characteristic common to all test data sets. Note that a lower RMS is expected when several grid lines are fitted to random velocities than when one H$_{\rm o}$ line alone is fitted, as can be seen by comparing figures 2 and 6. In no case, in these test data, was there a narrow RMS dip (width $\sim5$ H$_{\rm o}$-values) seen that had a depth below 0.8 of the RMS value of the adjacent baseline.

The broad RMS dip centered near H$_{\rm o}$ = 76 in Fig 7, is also visible in the FP data in Fig. 6 if the narrow dip near H$_{\rm o}$ = 71 is ignored as indicated by the dashed line. In the real data in Fig 6, however, the curve does not have the same symmetry visible for the uniformly distributed test curve in Fig. 7, and rises more steeply on the high-H$_{\rm o}$ side. This is discussed further below.

\subsection{Test 2}

In the second test we generated 10 random data sets containing 11 values (both positive and negative), each with a Gaussian distribution having the same dispersion as the real data and centered about a Hubble slope of 88. We again calculated RMS vs H$_{\rm o}$ curves for these data sets. These 10 curves were then averaged and the result is shown in Fig 8. This shows clearly that there was nothing in our data analysis that in any way favored the production of the quantized redshifts found previously by us and by others. Note that the resulting curve is again symmetrically spaced about an H$_{\rm o}$-value of 76.

\subsection{Test 3}

In our third test we used the real data and simply changed the \em polarity \em of the peculiar velocities derived assuming the Hubble slope of 88 in Fig 1, where approximately equal numbers of positive and negative velocities are seen. This approach had the advantage of producing a completely different velocity distribution while at the same time insuring that both the velocity dispersion and Hubble slope were identical to that of the real data. We again calculated the RMS deviation in V$_{\rm CMB}$ between the eleven sources and their nearest grid line, for a range of H$_{\rm o}$-values and z$_{f}$ = 0.62. The result is shown in Fig. 9. Here the solid curve was obtained using the original data for all eleven sources prior to changing the polarity of the "peculiar velocities". The dashed curve was obtained for the reversed polarity data. It is apparent from Fig. 9 that the reversed polarity curve is almost identical to that found for the real data, differing only in the presence of the deep RMS dip at H$_{\rm o}$ = 71 in the unadjusted real data. However, to obtain this result it was first necessary \em to rotate the dashed curve about the H$_{\rm o}$ = 76 line of symmetry found in the previous test curves.\em  Thus the following conclusions can be drawn. The asymmetry seen in the real data in Figs 6 and 9 (steeper slope on the high-H$_{\rm o}$ side), is not present when the source peculiar velocity distribution is symmetric (Gaussian or uniform) in Figs 7 and 8. Furthermore, it flips to the opposite side (steeper on the low-H$_{\rm o}$ side) when the polarity of the "peculiar velocities" is reversed. This indicates that \em the asymmetry is tied to the shape of the "peculiar velocity" distribution, and that the velocity distribution in the real data is therefore not symmetric about its center. \em This effect has been found to be even more pronounced in a follow-up analysis of spiral galaxies \citep{bel03}. It can be predicted if the "peculiar velocities" are really discrete intrinsic redshifts, whose density increases as the density of the grid lines increases at high values of $m$. This result is therefore seen as additional evidence favoring intrinsic redshifts.

Only for the real FP data was there a deep, narrow dip seen in the RMS. From the test data analysis we conclude that this dip in the RMS deviation in V$_{\rm CMB}$ in Fig 6, where the RMS value falls below the surrounding baseline by a factor of 5.8, would be \em unexpected \em if the intrinsic redshifts were random peculiar velocities. Although we cannot completely rule out the possibility that this RMS dip has occurred by chance, the test data suggests that the likelihood of this is small. This question can perhaps best be answered by examining a larger sample of galaxies for the presence of discrete redshifts. We have recently carried out a similar analysis on 55 spiral galaxies and have found that they too show evidence for the same intrinsic redshifts \citep{bel03}.

 For the RMS dip near H$_{\rm o}$ = 71 if Fig 6, the RMS deviation in V$_{\rm CMB}$ is reduced to $\sim 56$ km s$^{-1}$, confirming that there is an excellent fit to the z$_{\rm iG}$[1,m] intrinsic redshift lines in Fig. 5. Furthermore, the narrowness of this feature requires the simultaneous alignment of sources over a relatively large range of distances in Fig. 5. 

\begin{figure}
\hspace{-1.0cm}
\vspace{-1.0cm}
\epsscale{1.0}
\plotone{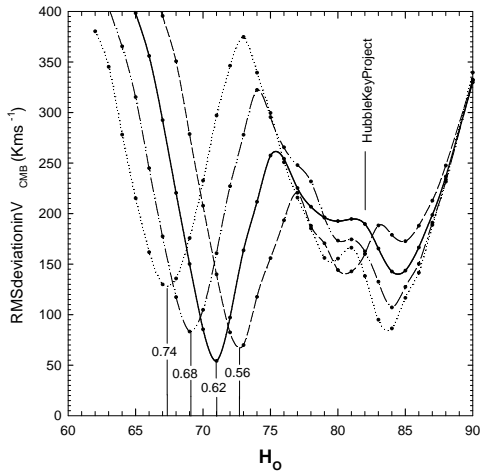}
\caption{\scriptsize{Plot of RMS deviation in V$_{\rm CMB}$ velocities relative to the grid lines as in Fig. 5, for 9 $N$ = 1 sources, as a function of the Hubble constant for several different z$_{f}$ values from 0.56 to 0.74. \label{fig10}}}
\end{figure}

\begin{figure}
\hspace{-1.0cm}
\vspace{-1.0cm}
\epsscale{1.0}
\plotone{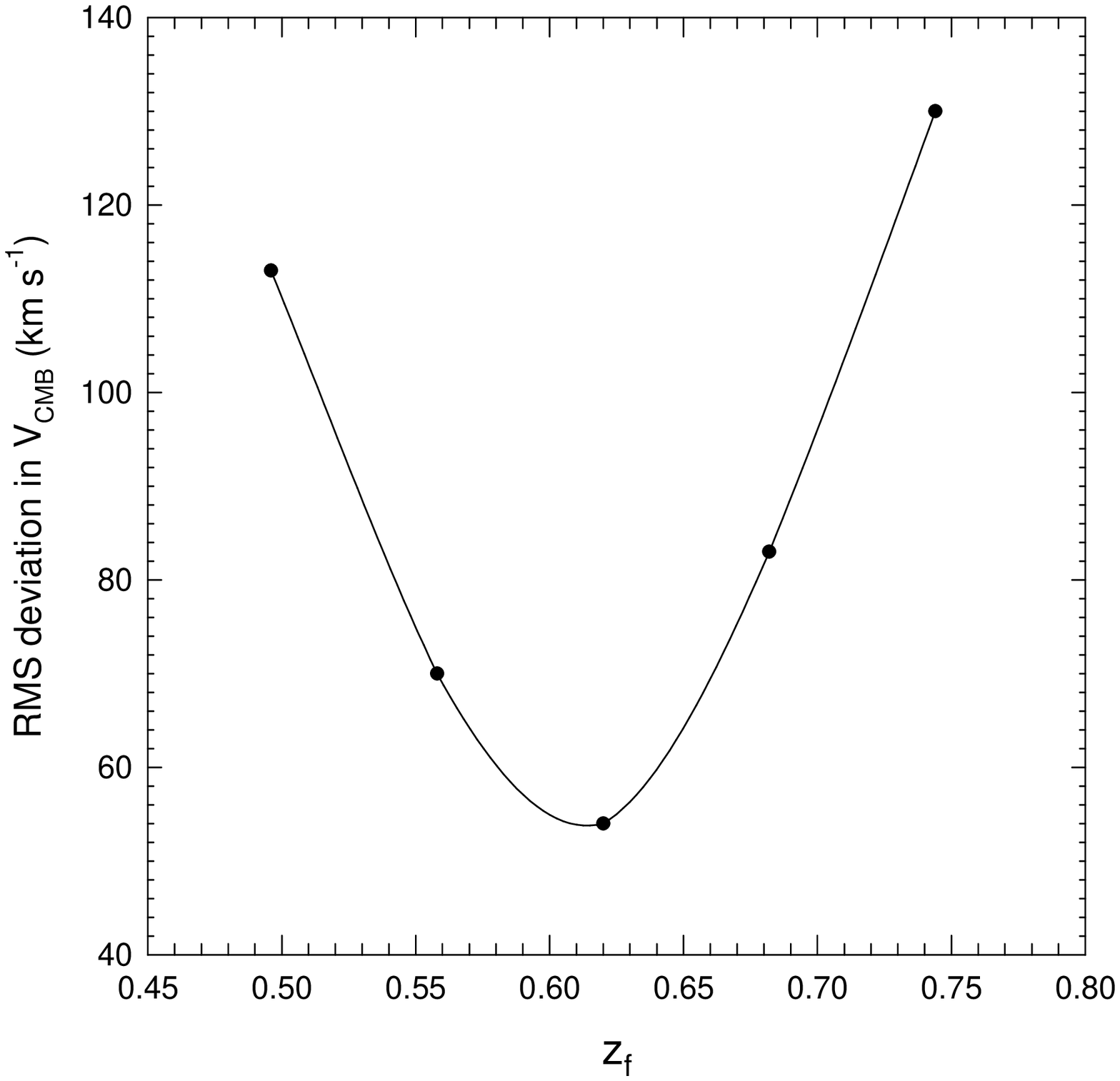}
\caption{\scriptsize{Plot of minimum RMS values in Fig. 10 vs z$_{f}$ for z$_{\rm iG}$[1,$m$] sources. \label{fig11}}}
\end{figure}

\section{An Independent Test To Determine the Value of z$_{f}$.}

Unlike the situation for the Hubble constant, where the uncertainty is still large, z$_{f}$ has been clearly defined on several previous occasions to be z$_{f}$ = $0.62 \pm0.01.$ Since this number determines the spacing between the intrinsic redshift grid lines in Fig. 4, we now have an opportunity to test previous results by finding the value of z$_{f}$ that gives the best fit to the data. To do this we used several z$_{f}$-values ranging from 0.56 to 0.94 and found that for the deep narrow feature in Fig. 6, both the $position$ and $depth$ of the minimum RMS value varied with z$_{f}$. Fig. 10 shows how the minimum value of the RMS deviation in V$_{\rm CMB}$ changes with z$_{f}$. The minimum values are plotted in Fig. 11 where the best fit occurs at z$_{f}$ = 0.613 $\pm 0.01$. This is interpreted as a confirmation of earlier work.

Although a second narrow feature is visible in the FP data near H$_{\rm o}$ = 84, at no time, as z$_{f}$ was varied, did its RMS value drop below one-half the value of the surrounding baseline. This can be compared to the RMS dip at H$_{\rm o}$ = 71 which drops by a factor of 5.8 below the surrounding baseline. Note that curves for z$_{f}$-values of 0.84 and 0.94 were also calculated but have not been included in Fig. 10 to avoid confusion. Note also that the intrinsic redshift grid pattern repeats every factor of 2 change in z$_{f}$ number.

The discrete velocities obtained for the FP clusters are listed in col 4 of Table 1. Col 5 lists the Hubble velocity of each cluster after removal of the intrinsic component.

From the above tests using randomly generated peculiar velocities, and from the fact that the best fit occurs for the previously predicted z$_{f}$-value of z$_{f}$ = 0.62$\pm0.01$, we conclude that the FP cluster redshifts very likely do contain some of the same discrete intrinsic redshifts found by \citet{tif97}, and their higher octave-spaced harmonics as defined by equation B1. This value (z$_{f}$ = 0.62$\pm0.01$) has been found previously in at least four independent investigations \citep{bur68,bur90,tif96,bel02c}. This result is therefore more than just a good fit of data points to a set of grid lines. The fact that the best fit is obtained at the predicted z$_{f}$ value makes it unlikely that this result has occurred by chance.

\begin{deluxetable}{ccccc}
\tabletypesize{\scriptsize}
\tablecaption{Parameters of FP Clusters. \label{tbl-1}}
\tablewidth{0pt}
\tablehead{
\colhead{Cluster/Group} & \colhead{D (Mpc)} & \colhead{V$_{\rm CMB}$ (km s$^{-1}$)\tablenotemark{a}} & \colhead{Transit.(Disc.Vel.)(km s$^{-1}$)\tablenotemark{b}} & \colhead{V$_{\rm H}$ (km s$^{-1}$)\tablenotemark{c}}
}

\startdata
Dorado     & 13.8  & 1131 & z$_{\rm iG}$[1,7](145.2)  & 986  \\ 
Grm 15     & 47.4  & 4530 & z$_{\rm iG}$[1,4](1157.9) & 3372  \\ 
Hydra      & 49.1  & 4061 & z$_{\rm iG}$[1,5](580.1)  & 3481  \\ 
Abell S753 & 49.7  & 4351 & z$_{\rm iG}$[2,6](725.2) & 3626  \\ 
Abell 3574 & 51.6  & 4749 & z$_{\rm iG}$[1,4](1157.9) & 3591  \\ 
Abell 194  & 55.9  & 5100 & z$_{\rm iG}$[1,4](1157.9) & 3942  \\ 
Abell S639 & 59.6  & 6533 & z$_{\rm iG}$[1,3](2314.3) &  4219  \\ 
Coma       & 85.8  & 7143 & z$_{\rm iG}$[1,4](1157.9) &  5985  \\ 
Abell 539  & 102.0 & 8792 & z$_{\rm iG}$[2,7](1448.6) & 7343  \\ 
DC 2345-28 & 102.1 & 8500 & z$_{\rm iG}$[1,4](1157.9) &  7342  \\ 
Abell 3381 & 129.8 & 11536 & z$_{\rm iG}$[1,3](2314.3) &  9222  \\

\enddata
\tablenotetext{a}{Total measured velocity including intrinsic component}
\tablenotetext{b}{intrinsic component obtained using z$_{f}$ = 0.62.}
\tablenotetext{c}{Hubble velocity after removal of discrete component}

\end{deluxetable}

\section{The Hubble Constant Between 10 and 110 Mpc}

In Fig. 12 the minimum RMS deviation in V$_{\rm CMB}$ for each different z$_{f}$ value examined is plotted vs H$_{\rm o}$. A quadratic regression fitted to these data gave a best-fit Hubble constant of 71.2 km s$^{-1}$ Mpc$^{-1}$. In Fig. 13 the Hubble flow velocities in column 5 of Table 1 have been plotted versus cluster distance. The solid line gives the result of a linear regression fit to all the FP data. The results of the fit are indicated in the figure and gave H$_{\rm o} = 71.4 \pm 0.6$ km s$^{-1}$ Mpc$^{-1}$, with a zero-velocity intercept of $-20 \pm 49$ km s$^{-1}$. The uncertainty in H$_{\rm o}$ listed in Fig. 13 is small, and as mentioned above indicates that the relative uncertainty in distance within the FP clusters also must be small (see below). However, the uncertainty in the absolute value of H$_{\rm o}$, which depends on the accuracy of calibrators, could be relatively large. Although the good RMS fit near H$_{\rm o}$ = 71 occurs at the same H$_{\rm o}$-value found for the remaining 4 groups examined by \citet{fre01}, this result is assumed here to be purely coincidental since its true uncertainty is likely to be similar to that reported by the Hubble Key Project. We conclude only that if the galaxy redshifts contain an intrinsic redshift component that is not taken into account, the Hubble constant obtained is likely to be 10-20 percent too large. This implies further that if intrinsic redshifts are present in other galaxies studied by the Hubble Key Project, the true value of the Hubble constant may be closer to H$_{\rm o}$ = 60. This is examined further in our follow-up paper on spiral galaxies \citep{bel03}.

\begin{figure}
\hspace{-1.0cm}
\vspace{-1.0cm}
\epsscale{1.0}
\plotone{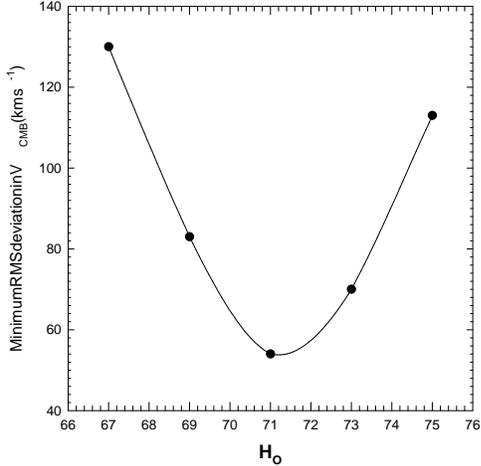}
\caption{\scriptsize{Plot of RMS minimum values in Fig. 7 vs Hubble constant for z$_{\rm iG}$[1,$m$] sources. \label{fig12}}}
\end{figure}

\begin{figure}
\hspace{-1.0cm}
\vspace{-1.0cm}
\epsscale{1.0}
\plotone{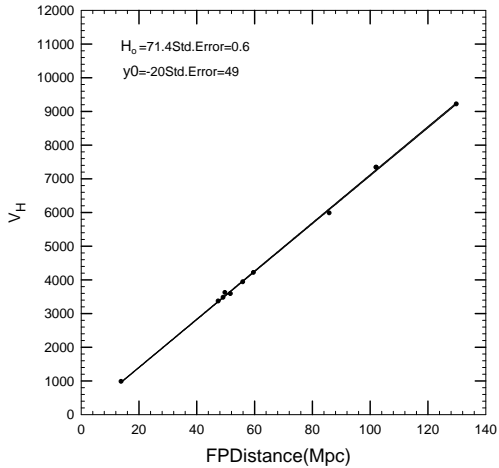}
\caption{\scriptsize{Plot of V$_{\rm H}$ from Table 1 (col 5) vs distance. The solid line represents the results of a linear regression fit to the data. \label{fig13}}}
\end{figure}

\subsection{Comments on the Uncertainties in Radial Velocity and Distance}

In this paper we have introduced the idea that most of the scatter in Hubble plots, previously assumed to be due to peculiar velocities or distance uncertainties, may instead be due largely to the presence of intrinsic redshift components. We assume that the distance uncertainties are approximately 1-2 percent. If so they will not prevent the detection of the discrete redshift components we are looking for. The redshift uncertainty introduced by the assumption that the redshift scatter is due to peculiar velocities is also largely removed when the intrinsic redshifts are identified and removed. A lower limit to the uncertainty in the radial velocities will be set by the velocity measurement error. The radial velocities for these clusters were measured by \citet{jor95} who state that comparisons of their radial velocities with data from the literature show that their determinations are accurate to within $\sim 35$ km s$^{-1}$. This will introduce a velocity scatter of this magnitude into the data. Since our high $m$ cutoff is at 145 km s$^{-1}$, and most of our discrete "velocities" are separated by several hundred km s$^{-1}$, this measurement error cannot prevent the detection of the discrete "velocities" we seek, if all large-scale motions have truly been accounted for. When taken together this suggests that the uncertainties in both velocity and distance are sufficiently good for our purpose. This question of distance uncertainties is examined in more detail in our follow-up paper using spiral galaxies \citep{bel03}.

\section{Discussion}

It is indeed remarkable that the very small discrete 'velocities' ($< 145$ km s$^{-1}$) found in galaxies have been detected in the presence of other Doppler components, and credit for this is due solely to the excellent work of W. Tifft. The fact that the discrete redshifts found in quasars have been shown to be tied directly to those found in galaxies \citep{bel02d}, when both were determined independently, suggests that they all have a common origin. However, it still leaves open the question of how to interpret them.

In addition to the scenario presented here and earlier \citep{bel02c,bel02d} that argues that all intrinsic redshifts in quasars and galaxies are harmonically related to the fundamental redshift z$_{f}$ = 0.62, two other possible scenarios have been discussed in the literature. The first of these is referred to here as the Karlsson model \citep{kar71,kar77,bur01}. It suggests that quasar redshifts are periodic in log(1+z). However, this description does not include the 0.06 period found by \citet{bur90} for quasars between z = 0.062 and z = 0.62. Nor does it include a link to the discrete velocities found in galaxies \citep[and this paper]{tif97,bel02d}.

A second interpretation has been proposed by \citet{tif96,tif02} which uses the fact that the discrete velocities in galaxies appear to be related to the constant $c$, the speed of light. In his model, referred to as the Lehto-Tifft model \citep{tif96,tif02}, the entire redshift is presumed to be quantized and to arise from time dependent decay from an origin at the Planck scale. In this model the decay process is a form of period doubling. However, as noted previously, Fig. 5 has significance for this model in that it shows clearly that the entire redshift is $not$ quantized. The quantized portion is superimposed on top of the Hubble flow. Although the \em discrete redshifts \em found in galaxies by \citet{tif96,tif97} have been confirmed here, the Lehto-Tifft $model$ is not compatible with the FP cluster data and is therefore effectively ruled out if the FP results in Fig. 5 are correct. The technique used by Tifft appears not to have allowed him to realize that the discrete redshifts were superimposed on top of the Hubble flow. Furthermore, his claim \citep{tif02} that his periods can be fitted to peaks in the Hubble Deep Field source distribution carries little weight when the number of periods he has to choose from is taken into account.

Fig. 5 also has significance for the evolution model proposed earlier \citep{bel02b,bel02c} in which it was suggested that galaxies are born throughout the entire age of the Universe as QSOs with large quantized, intrinsic redshifts that are harmonically related to z$_{f}$ = 0.62. These intrinsic redshifts decrease as the QSOs evolve into galaxies. The big bang is included in this model of the universe, and the intrinsic redshift components must then be superimposed on the Hubble flow, as confirmed here with the FP cluster data.

It is therefore concluded that the equations in Appendices A and B, that link all discrete redshifts to a fundamental redshift z$_{f}$ = 0.62, are currently the most complete for the purpose of defining the observed discrete redshifts.

\section{Conclusions}

We have identified discrete intrinsic redshifts in FP clusters that are identical to those predicted previously in galaxies \citep{bel02d,tif97}. When these are taken into account, we obtain a Hubble constant of H$_{\rm o}$ = 71.4$\pm0.6$ km s$^{-1}$ Mpc$^{-1}$, using all eleven FP clusters. However, we stress that this result does not necessarily mean that the true Hubble constant in the local Universe has been determined, and much larger source samples than used here will have to be studied first. When the discrete redshifts are removed, the RMS deviation in the Hubble velocities for the FP clusters is close to 56 km s$^{-1}$. This may be more in line with what is expected if all peculiar velocities have truly been taken into account, and all primordial motions have been damped out by adiabatic expansion. This small dispersion also implies that the relative uncertainties in distance within the FP clusters must be small. Finally, the FP cluster data have provided a new, and independent, way of determining z$_{f}$. Using it we obtain z$_{f}$ = 0.613 $\pm0.01$ which agrees with the value z$_{f}$ = 0.62 $\pm0.01$ determined independently in four previous investigations.

\acknowledgements

We thank D. McDiarmid, and J.K.G. Watson for helpful comments when this manuscript was being prepared.

\appendix
\section{Appendix A}

Intrinsic Redshifts in Quasars.

\citet{bel02c} found evidence that intrinsic redshifts in quasars can be defined by the relation:

\begin{equation}
z_{iQ}\:= z_{f}[N - 0.1M_{N}]
\end{equation}

where z$_{f}$ = 0.62, $N$ = 1,2,3,4,.. and

\begin{equation}
M_{1} = (n)_{n = 0,1,2,3,...9}
\end{equation}
\begin{equation}
M_{2} =  \left(\frac{n(n+1)}{2}\right)_{n = 0,1,2,3,4,5}
\end{equation}
\begin{equation}
M_{3} =  \left[\frac{n(n+1)}{2}\right]\left[\frac{\frac{n(n+1)}{2}+1}{2}\right]_{n = 0,1,2,3}
\end{equation}

If $N$ = 4 is not forbidden then:

\begin{equation}
M_{4} = \left(\frac{p(p+1)}{2}\right)_{n = 0,1,2}
\end{equation}
where
\begin{equation}
 p = \left[\frac{n(n+1)}{2}\right]\left[\frac{\frac{n(n+1)}{2}+1}{2}\right]
\end{equation}

For $N>4$, $n$ = 0 and 1 only, where z$_{\rm iQ}$[$N$,0] = 0.62$N$ and z$_{\rm iQ}$[$N$,1] = (z$_{\rm iQ}$[$N$,0] - 0.062).

Equation A1 represents a series of equations, one for each $N$ value. The first three of these are:

\begin{equation}
z_{iQ}[1,n]\:= z_{f}[1 - 0.1n]\;_{n = 0,1,2...9}
\end{equation}

\begin{equation}
z_{iQ}[2,n]\:= z_{f}\left[2 - 0.1\frac{n(n + 1)}{2}\right]\;_{n = 0,1,2,3,4,5}
\end{equation}

\begin{equation}
z_{iQ}[3,n]\:= z_{f}\left\{3 - 0.1\left( \frac{n(n+1)}{2} \right) \left(\frac{\frac{n(n+1)}{2} + 1}{2}\right) \right\}_{n = 0,1,2,3}
\end{equation}

Thus each quasar intrinsic redshift z$_{\rm iQ}$[$N,n$] is uniquely defined by the quantum numbers $N$ and $n$.

\section{Appendix B}

Intrinsic Redshifts in Galaxies

For galaxies the intrinsic redshift components are defined by the relation:

\begin{equation}
z_{iG}[N,m] = \left(\frac{z_{iQ}\left[N,n_{max}\right]}{2^{m}}\right)_{m=0,1,2,3..\infty}
\end{equation}

and each galaxy intrinsic redshift z$_{iG}$[$N$,$m$] is then uniquely defined by the quantum numbers $N$ and $m$. For $N$ = 1, $n_{\rm max}$ = 9 and z$_{\rm iQ}$[1,9] = 0.062. For $N$ = 2, $n_{\rm max}$ = 5 and z$_{iQ}$[2,5] = 0.310. For $N$ = 3, $n_{\rm max}$ = 3 and z$_{iQ}$[3,3] = 0.558. For $N$ = 4, $n_{\rm max}$ = 2, and z$_{iQ}$[4,2] = 1.178. For $N > 4, n_{\rm max}$ = 1 and z$_{iQ}$ = $N$z$_{f}$ - 0.062.


\clearpage

\begin{thebibliography}

\bibitem[Arp(2002)]{arp02} Arp, H. 2002, \apj, 571, 615
\bibitem[Bell(2002a)]{bel02a} Bell, M.B. 2002a, \apj, 566, 705
\bibitem[Bell(2002b)]{bel02b} Bell, M.B. 2002b, \apj, 567, 801 
\bibitem[Bell(2002c)]{bel02c} Bell, M.B. 2002c, astro-ph/0208320
\bibitem[Bell(2002d)]{bel02d} Bell, M.B. 2002d, astro-ph/0211091
\bibitem[Bell and Comeau(2003)]{bel03} Bell, M.B., and Comeau, S.P. 2003, (ApJ, submitted)
\bibitem[Burbidge(1968)]{bur68} Burbidge, G. 1968, \apj, 154, L41
\bibitem[Burbidge and Hewitt(1990)]{bur90} Burbidge, G. and Hewitt, A. 1990, \apj, 359, L33
\bibitem[Burbidge and Napier (2001)]{bur01} Burbidge, G. and Napier, W.M. 2001, 
\bibitem[Freedman et al.(2001)]{fre01} Freedman, W.L., Madore, B.F., Gibson, B.K., Ferrarese, L, Kelson, D.D., Sakai, S., Mould, J.R., Kennicutt, R.C., Ford, H.C., Graham, J.A, Huchra, J.P., Hughes, S.M.G., Illingworth, G.D., Macri, L.M., and Stetson, P.B. 2001, \apj, 553, 47
\bibitem[Jorgensen et al.(1995)]{jor95} Jorgensen, I., Franx, M., and Kjaergaard, P. 1995, \mnras, 276, 1341
\bibitem[Jorgensen et al.(1996)]{jor96} Jorgensen, I., Franx, M., and Kjaergaard, P. 1996, \mnras, 280, 167
\bibitem[Karlsson (1971)]{kar71} Karlsson, K.G. 1971, \aap, 13, 333
\bibitem[Karlsson (1977)]{kar77} Karlsson, K.G. 1977, \aap, 58, 237
\bibitem[Kraan-Korteweg(1986)]{kra86} Kraan-Korteweg, R.C. 1986, \aap, 66, 255
\bibitem[Russell(2002)]{rus02} Russell, D. 2002, \apj, 565, 681
\bibitem[Tifft(1996)]{tif96} Tifft, W.G. 1996, \apj, 468, 491
\bibitem[Tifft(1997)]{tif97} Tifft, W.G. 1997, \apj, 485, 465
\bibitem[Tifft(2002)]{tif02} Tifft, W.G. 2002, \apss (in press)
\end{thebibliography}
\end{document}